\documentstyle[12pt]{article}
\input tcilatex

\begin{document}

\title{{\bf Degenerate and Quasi Degenerate Majorana Neutrinos}}
\author{G. C. Branco$^a$ \thanks{%
E-mail address : d2003@beta.ist.utl.pt} , M. N.\ Rebelo$^a$ \thanks{%
E-mail address : rebelo@beta.ist.utl.pt} and J. I. Silva-Marcos$^b$ \thanks{%
E-mail address : juca@nikhef.nl} \and \and
$^a${\it Centro de F\'\i sica das Interac\c c\~oes Fundamentais, CFIF,} \\
{\it Instituto Superior T\'ecnico,}\\
{\it Av. Rovisco Pais, P-1096 Lisboa-Codex, Portugal }\\
$^b${\it NIKHEF, Kruislaan 409, 1098 SJ Amsterdam, The Netherlands}}
\date{}
\maketitle

\begin{abstract}
We study mixing and CP violation of three left-handed Majorana neutrinos in
the limit of exactly degenerate masses, identify the weak-basis invariant
relevant for CP violation and show that the leptonic mixing matrix is
parametrized only by two angles and one phase. After the lifting of the
degeneracy, this parametrization accommodates the present data on
atmospheric and solar neutrinos, as well as double $\beta $ decay. Some of
the leptonic mixing ans\"atze suggested in the literature correspond to
special cases of this parametrization.
\end{abstract}

\setlength{\baselineskip}{14pt} 
\renewcommand{\thesubsection}{\arabic{subsection}} 
\begin{picture}(0,0)
       \put(335,420){FISIST/11-98/CFIF}
\end{picture}
\vspace{-24pt} \thispagestyle{empty}



\newpage

The Superkamiokande collaboration $\cite{ref1}$ has recently provided
evidence confirming the atmospheric neutrino anomaly, as well as the solar
neutrino deficit. The interpretation of these experimental results within
the framework of three left-handed neutrinos, without sterile neutrinos,
together with the assumption that relic neutrinos constitute the hot dark
matter of the universe $\cite{ref1a}$, inescapably leads to highly
degenerate neutrinos $\cite{ref2}$.

In this letter, we analyse in detail neutrino mixing and CP violation in the
case of three Majorana neutrinos with exactly degenerate masses and then
consider the case of quasi degenerate masses. We identify the weak-basis
invariant, which controls the strength of CP violation in the limit of exact
mass degeneracy, and point out that, in this limit, the neutrino mixing
matrix is in general parametrized by two angles and one phase. We then show
that a two-angle parametrization suggested by the exact degeneracy limit can
fit all the present atmospheric and solar neutrino data, and complies with
the bound imposed by neutrinoless beta decay. Furthermore, we point out that
various of the recently suggested neutrino mixing schemes, such as the
bimaximal mixing $\cite{ref3}$, the democratic mixing $\cite{ref4}$, as well
as the scheme suggested by Georgi and Glashow $\cite{ref5}$, correspond to
specific cases of our two-angle parametrization.

{\bf The limit of exact degeneracy}. Let us consider three left-handed
neutrinos and introduce a generic Majorana mass term, 
\begin{equation}
\label{eq1}{\cal L}_{{{\rm {mass}}}}\ =\ -\ (\nu _{L_\alpha })^T\ C^{-1}\
m_{\alpha \beta \quad }\nu _{L_\beta }\ +{\rm {\ h.c.}} 
\end{equation}
where $m=(m_{\alpha \beta })$ is a $3\times 3$ complex symmetric mass
matrix, and $\nu _{L_\alpha }$ denote the left-handed weak eigenstates. We
shall work in the weak-basis (WB) where the charged lepton mass matrix is
diagonal, real and positive. The neutrino mass matrix can be diagonalized by
the transformation \footnote{%
The neutrino mass matrix $m$ could be an effective Majorana mass matrix
within a framework with three left-handed and three right-handed neutrinos.},

\begin{equation}
\label{eq2}U^T\cdot m\cdot U\ =\ {{\rm {diag}}(m_{\nu _1},\ m_{\nu _2},\
m_{\nu _3})} 
\end{equation}
The weak eigenstates, $\nu _{L_\alpha }$, are related to the mass
eigenstates, $\nu _{L_i}$, by $\nu _{L_\alpha }=U_{\alpha i}$ $\nu _{L_i}$,
so that the charged current of the lepton weak interactions is given by:

\begin{equation}
\label{eq3}{\cal L}_{{{\rm {W}}}}\ =\ \frac g2\ \ \left( 
\begin{array}{ccc}
\overline{e}, & \overline{\mu }, & \overline{\tau } 
\end{array}
\right) _L\gamma _\mu \quad U\quad \left( 
\begin{array}{c}
\nu _1 \\ 
\nu _2 \\ 
\nu _3 
\end{array}
\right) _LW^\mu \quad +{\rm {\ h.c.}} 
\end{equation}

It is well known that, for the non-degenerate case, the neutrino
diagonalization matrix $U$ can be parametrized by three angles and three
phases that are CP violating. In the limit of exact degeneracy, we shall
show here that, in general, $U$ can not be rotated away, and its
parametrization requires two angles and one CP violating phase. Furthermore,
we shall see that only if the theory is CP invariant and the three
degenerate neutrinos have the same CP parity, can the matrix $U$ be rotated
away. This is to be contrasted to the case of Dirac neutrinos, where there
is no mixing or CP violation in the exact degeneracy limit.

Let us consider the limit of exact degeneracy with $\mu $ the common
neutrino mass. It is useful to define the dimensionless matrix $Z_{\circ }=\
m/\mu $ which from Eq.(\ref{eq2}) can be written as:

\begin{equation}
\label{eq4}Z_{\circ }\ =U_{\circ }^{\star }\cdot U_{\circ }^{\dagger }\ 
\end{equation}
where $U_{\circ }$ denotes the mixing matrix in the exact degeneracy limit.
It follows from Eq.(\ref{eq4}) that $Z_{\circ }$ is a unitary symmetric
matrix.\ By making a WB transformation, under which $Z_{\circ }\rightarrow \
K\cdot Z_{\circ }\cdot K$, with $K$ a diagonal unitary matrix, it is
possible to choose the first line and the first column of $Z_{\circ }$ real,
while keeping the charged lepton mass matrix diagonal real and positive.
Without loss of generality, the matrix $Z_{\circ }$ can then be written as: 
\begin{equation}
\label{eq5}Z_{\circ }\ =\ \left( 
\begin{array}{ccc}
1 & 0 & 0 \\ 
0 & c_\phi & s_\phi \\ 
0 & s_\phi & -c_\phi 
\end{array}
\right) \cdot \left( 
\begin{array}{ccc}
c_\theta & s_\theta & 0 \\ 
s_\theta & z_{22} & z_{23} \\ 
0 & z_{23} & z_{33} 
\end{array}
\right) \cdot \left( 
\begin{array}{ccc}
1 & 0 & 0 \\ 
0 & c_\phi & s_\phi \\ 
0 & s_\phi & -c_\phi 
\end{array}
\right) 
\end{equation}
Unitarity of $Z_{\circ }$, then implies that either $s_\theta $ or $z_{23}$
must vanish. It can be readily verified that the case $s_\theta =0$
automatically leads to CP invariance. Assuming $s_\theta \neq 0$, then the
most general form for the symmetric unitary matrix $Z_{\circ }$ is given by: 
\begin{equation}
\label{eq6}Z_{\circ }\ =\ \left( 
\begin{array}{ccc}
1 & 0 & 0 \\ 
0 & c_\phi & s_\phi \\ 
0 & s_\phi & -c_\phi 
\end{array}
\right) \cdot \left( 
\begin{array}{ccc}
c_\theta & s_\theta & 0 \\ 
s_\theta & -c_\theta & 0 \\ 
0 & 0 & e^{i\alpha } 
\end{array}
\right) \cdot \left( 
\begin{array}{ccc}
1 & 0 & 0 \\ 
0 & c_\phi & s_\phi \\ 
0 & s_\phi & -c_\phi 
\end{array}
\right) 
\end{equation}

The parametrization of $Z_{\circ }$ in Eq.(\ref{eq6}) does not include the
trivial case where CP is a good symmetry and all neutrinos have the same CP
parity. In order to show that this is indeed the case, let us assume that
the most general $Z_{\circ }$ for degenerate neutrinos given by Eq.(\ref{eq4}%
), is a real matrix, so that CP invariance holds. In that case, $Z_{\circ }$
can be diagonalized by an orthogonal transformation $Z_{\circ }\rightarrow \
O^T\cdot Z_{\circ }\cdot O$, which leaves invariant both ${\rm {Tr}}%
(Z_{\circ })$ and $\det (Z_{\circ })$. Apart from trivial permutations, the
eigenvalues of $Z_{\circ }$ will be $(1,1,1)$, $(1,-1,1)$ or $(1,-1,-1)$. It
is well known $\cite{ref6}$ that the first case corresponds to three
neutrinos with the same CP parity, while the other two cases correspond to
one of the neutrinos having a CP parity opposite to that of the other two.
Now, in the parametrization of Eq.(\ref{eq6}), one obtains $\det (Z_{\circ
})=-e^{i\alpha }$, ${\rm {Tr}}(Z_{\circ })=e^{i\alpha }$, and therefore the
cases $(1,-1,1)$ and $(1,-1,-1)$ can be obtained, corresponding to $\alpha
=0 $ and $\alpha =\pi $, respectively. Obviously the case $(1,1,1)$ cannot
be obtained by the parametrization of Eq.(\ref{eq6}). As we previously
mentioned, this case corresponds to a trivial mixing matrix, which can be
rotated away. The matrix $Z_{\circ }$ given by Eq.(\ref{eq6}) can be
diagonalized through the transformation of Eq.(\ref{eq2}), with $U_{\circ }$
given by 
\begin{equation}
\label{eq7}U_{\circ }\ =\ \left( 
\begin{array}{ccc}
1 & 0 & 0 \\ 
0 & c_\phi & s_\phi \\ 
0 & s_\phi & -c_\phi 
\end{array}
\right) \cdot \left( 
\begin{array}{ccc}
\cos (\frac \theta 2) & \sin (\frac \theta 2) & 0 \\ 
\sin (\frac \theta 2) & -\cos (\frac \theta 2) & 0 \\ 
0 & 0 & e^{-i\alpha /2} 
\end{array}
\right) \cdot \left( 
\begin{array}{ccc}
1 & 0 & 0 \\ 
0 & i & 0 \\ 
0 & 0 & 1 
\end{array}
\right) 
\end{equation}
The matrix $U_{\circ }$ is then the mixing matrix appearing in the leptonic
charged currents. Given the Majorana character of neutrino masses and the
fact that $U_{\circ }$ is not an orthogonal matrix, it is clear that one can
not rotate away $U_{\circ }$ through a redefinition of the neutrino fields.
This is the case even in the CP invariance limit, i.e., $\alpha =0$, $\pi $.

{\bf The strength of CP violation and a WB invariant}. We have seen that CP
violation may arise even when the three Majorana neutrinos have identical
mass $\cite{ref7}$. Now, we present a weak-basis invariant which controls
the strength of the CP violation in the limit of exact degeneracy. It can be
readily verified that a necessary and sufficient condition for CP
invariance, in the degenerate limit, is: 
\begin{equation}
\label{eq8}G\equiv \ {\rm {Tr}}\left[ \ (m\cdot h\cdot m^{\star })\ ,\
h^{\star }\right] ^3\ =\ 0
\end{equation}
where $h=m_\ell \cdot m_\ell ^{\dagger }$, and $m_\ell $ denotes the charged
lepton mass matrix. The non-vanishing of $G$ signals CP violation, while the
vanishing of $G$ implies CP invariance in the limit of mass degeneracy.
Since $G$ is a WB invariant, it can be expressed in terms of lepton masses
and mixings. In the evaluation of $G$, it is convenient to choose the WB
where $h$ is diagonal, i.e., $h=$diag$\ (m_e^2,\ m_\mu ^2,\ m_\tau ^2)$. One
obtains: 
\begin{equation}
\label{eq9}G=6i\ \Delta _m~{\rm {Im}}[(Z_{\circ })_{11}(Z_{\circ
})_{22}(Z_{\circ })_{12}^{\star }(Z_{\circ })_{21}^{\star }]=\frac{3i}2\
\Delta _m~\cos (\theta )\sin ^2(\theta )\ \sin ^2(2\phi )\ \sin (\alpha )
\end{equation}
where $\Delta _m=$\ $\mu ^6\ (m_\tau ^2-m_\mu ^2\ )^2(m_\tau ^2-m_e^2\
)^2(m_\mu ^2-m_e^2\ )^2$ is a multiplicative factor which contains the
different masses of the charged leptons and the common neutrino mass $\mu $.
In Ref.$\cite{ref7}$ various examples of CP-odd WB-invariants were
constructed, but all of those invariants automatically vanish in the limit
of exact degeneracy. The special feature of the WB-invariant of Eq.(\ref{eq8}%
) is the fact that, in general, it does not vanish, even in the limit of
exact degeneracy of the three Majorana neutrino masses.

Since in the limit of exact degeneracy there is only one independent WB
invariant controlling the strength of CP violation, it is meaningful to ask
when is CP violation maximal. From Eq.(\ref{eq9}), it follows that $G$
assumes its maximal value for $\phi =\pi /4$, $\alpha =\pi /2$ and $\sin
(\theta )=\sqrt{2}/\sqrt{3}$, $\cos (\theta )=1/\sqrt{3}$. For these values
of $\phi $, $\theta $, $\alpha $ the matrix $Z_{\circ }$ assumes a very
special form: 
\begin{equation}
\label{eq9a}Z_{\circ }\ =K\cdot \frac 1{\sqrt{3}}\ \left( 
\begin{array}{ccc}
\omega & 1 & 1 \\ 
1 & \omega & 1 \\ 
1 & 1 & \omega 
\end{array}
\right) \cdot K 
\end{equation}
with $\omega =e^{-i2\pi /3}$ and $K=~{\rm {diag}}(e^{i\pi /3},e^{-i\pi
/3},e^{-i\pi /3})$. Thus the imposition of maximal CP violation leads to a
structure of the Majorana neutrino mass of the type that one obtains in the
framework of universal strength for Yukawa couplings $\cite{ref7a}$.

{\bf Lifting the degeneracy }. We have seen that, in the limit of exact
degeneracy, the leptonic mixing matrix can be parametrized by two angles $%
\theta $, $\phi $ and one phase $\alpha $. Obviously, the physically
interesting case corresponds to quasi-degenerate neutrinos. The degeneracy
is lifted through a small pertubation: 
\begin{equation}
\label{eq10}Z=Z_{\circ }+\varepsilon \ Q\ \ 
\end{equation}
where $\varepsilon $ is a small parameter and $Q$ is a symmetric complex
matrix of order one. At this stage, it is worth recalling that in the exact
degeneracy limit, the neutrino mixing matrix $U_{\circ }$ is only defined up
to an arbitrary orthogonal transformation $U_{\circ }\rightarrow U_{\circ
}\cdot O$. In the presence of a small pertubation $\varepsilon \ Q$, the
full matrix $Z$ will be diagonalized by a matrix $U=(U_{\circ }\cdot O)\cdot
W$, where $W$ is a unitary matrix close to the identity. In first order we
have: 
\begin{equation}
\label{eq11a}W={1\>\!\!\!{\rm I}}+i\varepsilon \ P
\end{equation}
with $P$ an hermitian matrix. In view of the above, it is useful to
diagonalize $Z$ in two steps. First, we make the transformation, 
\begin{equation}
\label{eq11b}Z\rightarrow Z^{\prime }\equiv U_{\circ }^T\cdot Z\cdot
U_{\circ }={1\>\!\!\!{\rm I}}+\varepsilon \ Q^{\prime }
\end{equation}
where we have used the fact that $U_{\circ }^T\cdot Z_{\circ }\cdot U_{\circ
}={1\>\!\!\!{\rm I}}$, and have defined $Q^{\prime }=U_{\circ }^T\cdot
Q\cdot U_{\circ }$. The matrix $Z^{\prime }$ is then diagonalized by, 
\begin{equation}
\label{eq11c}Z^{\prime }\rightarrow (OW)^T\cdot Z^{\prime }\cdot (OW)={%
1\>\!\!\!{\rm I}}+\varepsilon \ d
\end{equation}
where $d$ is diagonal and real. Using Eqs.(\ref{eq11a}), (\ref{eq11b}) and (%
\ref{eq11c}) one obtains in leading order of the pertubation: 
\begin{equation}
\label{eq12}O^T\cdot A\cdot O=d\quad ;\quad P+P^T=-O^T\cdot B\cdot O
\end{equation}
where $A$, $B$ are real symmetric matrices defined by $A={\rm {Re}}%
(Q^{\prime })$, $B={\rm {Im}}(Q^{\prime })$. Eqs.(\ref{eq12}) have a simple
interpretation. In the presence of a small pertubation around the degeneracy
limit, the mixing matrix becomes, to leading order, $U_{\circ }\cdot O$,
where $O\,$ is no longer arbitrary, being the orthogonal matrix which
diagonalizes the symmetric real matrix $A$. We have, of course, assumed that
the degeneracy is lifted in first order of pertubation. From the above
discussion it is clear that for quasi-degenerate neutrinos, in leading
order, only one CP violating phase appears in the leptonic mixing matrix,
namely the phase $\alpha $ present in $U_{\circ }$.

{\bf Phenomenological implications}. At this stage, one may ask whether,
after the lifting of the degeneracy, the two-angle parametrization given by
Eq.(\ref{eq7}) can still accommodate the present experimental data on
atmospheric and solar neutrinos, as well as the constraints on double beta
decay. It will be shown that this is indeed the case and, in fact, some of
the ans\"atze suggested in the literature are special cases of this
parametrization.

{\bf Double beta decay}. Let us first consider the constraints arising from
neutrinoless double beta decay, which can only occur if neutrinos are of
Majorana type, irrespective of whether or not there is CP violation or
non-trivial neutrino mixing. The amplitude for neutrinoless double beta
decay is proportional to $<m>$, an average neutrino mass, given in standard
notation by: 
\begin{equation}
\label{eq13}<m>\ =\ {\sum}_i\ U_{ei}^2\ m_{\nu _i}\ =\ m_{ee}^{\star }\ \ 
\end{equation}
where the $U_{ei}$ denote the elements of the first row of the mixing matrix 
$U$, and $m_{ee}$ is the $(1,1)$ element of the mass matrix $m$. The
experimental upper bound on $<m>$ depends on the model that is used for the
nuclear matrix elements. At present, the strongest bound is $%
|<m>|=|m_{ee}|<0.46\ eV$ $\cite{ref8}$. In the limit of exact degeneracy, we
have $m_{ee}=\mu \ \cos (\theta )$, where we have used the parametrization
of Eq.(\ref{eq6}). If we fix $\mu =2\ eV$, then neutrino masses are equal to
a precision sufficient to neglect their differences, and the experimental
bound on $m_{ee}$ immediately translates into a single bound on the
parameter $\theta $, namely $|\cos (\theta )|<0.23$.

{\bf Atmospheric and Solar Neutrino Data}. The atmospheric neutrino data
supports the existence of oscillations of atmospheric neutrinos to tau
neutrinos or to a sterile neutrino, with a large mixing angle satisfying the
bound $\sin ^2(2\theta _{{\rm {atm}}})>0.82$, and the neutrino mass square
difference in the range $5\times 10^{-4}\ eV^2<\Delta m_{{\rm {atm}}%
}^2<6\times 10^{-3}\ eV^2$. Recent data from the CHOOZ collaboration $\cite
{ref9}$ provides on the other hand some evidence against the possibility
that atmospheric muon neutrinos oscillate into electron neutrinos, although
in some special scenarios this possibility might still be open $\cite{ref10}$%
.

In the context of three left-handed neutrinos, the probability for a
neutrino $\nu _\alpha $ to oscillate to other neutrinos is: 
\begin{equation}
\label{eq13a}1-P(\nu _\alpha \rightarrow \nu _\alpha )=4\ {\sum }_{i<j}\
U_{\alpha i}U_{\alpha i}^{\star }U_{\alpha j}^{\star }U_{\alpha j}\quad \sin
^2\left[ \frac{\Delta m_{ji}^2}4\frac LE\right] 
\end{equation}
where $\Delta m_{ji}^2=|m_j^2-m_i^2|$, $E$ is the neutrino energy and $L$
denotes the distance travelled by the neutrino between the source and the
detector. Since in the range $L/E$ that is relevant for atmospheric
neutrinos the term in $\sin ^2[(\Delta m_{21}^2/4)(L/E)]$ can be
disregarded, we may identify $\sin ^2(2\theta _{{\rm {atm}}})$ with $%
4(U_{21}U_{21}^{\star }U_{23}^{\star }U_{23}+$ $U_{22}U_{22}^{\star
}U_{23}^{\star }U_{23})$.\ In the framework of our two-angle parametrization
of Eq.(\ref{eq7}), the above combination of matrix elements has a simple
form and one obtains $\sin ^2(2\theta _{{\rm {atm}}})=\sin ^2(2\phi )$,
i.e., $\theta _{{\rm {atm}}}$ can be identified with the angle $\phi $ and
thus the atmospheric neutrino data leads to the constraint $\sin ^2(2\phi
)>0.82$.

The discrepancy between the observed and the calculated $\cite{ref11}$ solar
neutrino fluxes also requires neutrino oscillations, although at this stage
various schemes are still possible, namely within the framework of the MSW
mechanism $\cite{ref12}$ there is a small angle solution $\sin ^2(2\theta _{%
{\rm {sol}}})\approx 7\times 10^{-3}\ $ with $\Delta m_{{\rm {sol}}%
}^2\approx 6\times 10^{-6}\ eV^2$,  and a large angle solution $\sin
^2(2\theta _{{\rm {sol}}})\sim 0.6-0.8$ with $\Delta m_{{\rm {sol}}%
}^2\approx 9\times 10^{-6}\ eV^2$. Another solution could be vacuum
oscillations with $\sin ^2(2\theta _{{\rm {sol}}})\approx 0.9$ and  $\Delta
m_{{\rm {sol}}}^2\approx 10^{-10}\ eV^2$. Since in our two-angle
parametrization one has $U_{13}=0$ we obtain $\sin ^2(2\theta _{{\rm {sol}}%
})=$ $4U_{11}U_{11}^{\star }U_{12}^{\star }U_{12}$ leading to $\sin
^2(2\theta _{{\rm {sol}}})=\sin ^2(\theta )$, i.e., in our parametrization $%
2\theta _{{\rm {sol}}}=\theta $.

From the above analysis it follows, that an attractive feature of this
two-angle parametrization is the fact that each of the experiments
considered, independently constrains a single parameter: double beta decay
and solar neutrino data only constrain $\theta $, while atmospheric neutrino
data only put a bound on $\phi $.

There have been several attempts to fit solar and atmospheric neutrino data.
The form of the matrix $U$ strongly depends on the scheme adopted to explain
the solar puzzle, with large or small mixing. It is clear that with small
mixing, no strong cancellation in the summation in Eq.(\ref{eq13}) can
occur, so in this case the double beta decay would forbid quasi-degenerate
neutrinos with masses in the range of cosmological relevance.

Next, we show that some of the neutrino mixing schemes proposed in the
literature correspond to specific cases of the two-angle parametrization
suggested by Eq.(\ref{eq7}).

(a) {\it Bimaximal Mixing} $\cite{ref3}$: In this scheme the lines of the
neutrino mixing matrix have the following structure: 
\begin{equation}
\label{eq14}
\begin{array}{ccc}
L_1=\left( \frac 1{\sqrt{2}},\frac{-1}{\sqrt{2}},0\right) ;\quad & 
L_2=\left( \frac 12,\frac 12,\frac 1{\sqrt{2}}\right) ;\quad & L_3=\left( 
\frac{-1}2,\frac{-1}2,\frac 1{\sqrt{2}}\right) 
\end{array}
\end{equation}
This pattern of neutrino mixing is obtained within the two-angle
parametrization for the following values of $\theta $, $\phi $ and $\alpha $%
: 
\begin{equation}
\label{eq15}\alpha =0\ ;\quad \cos (\theta /2)=-\sin (\theta /2)=-\cos (\phi
)=\sin (\phi )=\frac 1{\sqrt{2}} 
\end{equation}

(b) {\it Democratic Mixing} $\cite{ref4}$: This mixing has been proposed
within the framework of a ``d{emocratic'' structure for the quark and lepton
mass matrices. It was pointed out $\cite{ref4}$ that this neutrino mixing
automatically arises if one assumes that in the exact democratic limit,
neutrinos have no mass, and only acquire mass through diagonal
democracy-breaking terms. In this case the neutrino mixing matrix has, to a
very good approximation, the following form: 
\begin{equation}
\label{eq16}
\begin{array}{ccc}
L_1=\left( \frac 1{\sqrt{2}},\frac{-1}{\sqrt{2}},0\right) ;\quad & 
L_2=\left( \frac 1{\sqrt{6}},\frac 1{\sqrt{6}},\frac{-2}{\sqrt{6}}\right)
;\quad & L_3=\left( \frac 1{\sqrt{3}},\frac 1{\sqrt{3}},\frac 1{\sqrt{3}%
}\right) 
\end{array}
\end{equation}
Within the two-angle parametrization, one obtains the democratic mixing for
the following values of the parameters: 
\begin{equation}
\label{eq17}\alpha =0\ ;\quad \cos (\theta /2)=-\sin (\theta /2)=\frac 1{%
\sqrt{2}}\ ;\quad \cos (\phi )=\frac 1{\sqrt{2}}\sin (\phi )=\frac{-1}{\sqrt{%
3}} 
\end{equation}
}

In the above analysis, we have not paid attention to the factors ``$i$''
appearing in our two-angle parametrization of Eq.(\ref{eq7}). As we have
previously emphasized, these factors of ``$i$'' have to do with the fact
that in the construction of the two-angle parametrization, we have
implicitly assumed that in the limit of CP\ invariance (i.e. $\sin (\alpha
)\rightarrow 0$), one of the Majorana neutrinos has relative CP parity
opposite to the other two. The factors of ``$i$'' do not play any r\^ole in
the analysis of atmospheric and solar neutrino data, but are crucial in the
analysis of double beta decay.

(c) {\it Georgi-Glashow mass matrix} $\cite{ref5}$: Using an analysis of the
present neutrino data Georgi and Glashow have suggested the following
approximate form for the Majorana neutrino mass matrix 
\begin{equation}
\label{eq18}
\begin{array}{ccc}
(m)_{1i}=\mu \left( 0,\frac 1{\sqrt{2}},\frac 1{\sqrt{2}}\right) ;\quad & 
(m)_{2i}=\mu \left( \frac 1{\sqrt{2}},\frac 12,\frac{-1}2\right) ;\quad & 
(m)_{3i}=\mu \left( \frac 1{\sqrt{2}},\frac{-1}2,\frac 12\right) 
\end{array}
\end{equation}
From Eq.(\ref{eq6}) it follows that this neutrino mass matrix is obtained,
within the two-angle parametrization for the following values of its
parameters, 
\begin{equation}
\label{eq19}
\begin{array}{ccc}
\alpha =0; & \sin (\theta )=1; & \cos (\phi )=\sin (\phi )=\frac 1{\sqrt{2}} 
\end{array}
\end{equation}

To summarize, we have built a general parametrization for the leptonic
mixing matrix in the case of three exactly degenerate Majorana neutrinos,
caracterized by two angles and one phase and have shown that for
quasi-degenerate neutrinos, this parametrization accommodates all present
neutrino data. A remarkable feature of this parametrization is the fact that
each of the relevant experiments considered (solar, atmospheric and double
beta decay) independently constrains a single angle. We have also presented
a weak-basis invariant which controls the strength of CP violation in the
case of exact degeneracy.

Acknowledgment: This work was partially supported by the ``Funda\c c\~ao
para a Ci\^encia e Tecnologia'' of the Portuguese Ministry of Science and
Technology.

\end{document}